\documentclass{ieeetj}
\usepackage{cite}
\usepackage{amsmath,amssymb,amsfonts,bm, mathtools,multirow}
\usepackage{algorithmic}
\usepackage{graphicx,color}
\usepackage{textcomp}
\usepackage{xcolor}
\usepackage{hyperref}
\let\labelindent\relax
\usepackage{enumitem}
\newtheorem{definition}{Definition}
\hypersetup{hidelinks=true}
\usepackage{algorithm,algorithmic}
\usepackage{booktabs}
\def\BibTeX{{\rm B\kern-.05em{\sc i\kern-.025em b}\kern-.08em
    T\kern-.1667em\lower.7ex\hbox{E}\kern-.125emX}}
\AtBeginDocument{\definecolor{tmlcncolor}{cmyk}{0.93,0.59,0.15,0.02}\definecolor{NavyBlue}{RGB}{0,86,125}}

\def\OJlogo{\vspace{-4pt}$<$Society logo(s) and publication title will appear here.$>$}
\def\seclogo{\vspace{10pt}$<$Society logo(s) and publication title will appear here.$>$}

\def\authorrefmark#1{\ensuremath{^{\textbf{#1}}}}

\begin{document}
\receiveddate{XX Month, XXXX}
\reviseddate{XX Month, XXXX}
\accepteddate{XX Month, XXXX}
\publisheddate{XX Month, XXXX}
\currentdate{XX Month, XXXX}
\doiinfo{XXXX.2022.1234567}

\markboth{}{Author {et al.}}

\title{Dynamic Sensor Scheduling Based on Node Partitioning of Graphs}

\author{Ryosuke Ikura\authorrefmark{1}, Junya Hara \authorrefmark{1} (Member, IEEE), Hiroshi Higashi \textbf{\authorrefmark{1}} (Member, IEEE),\\ and Yuichi Tanaka\authorrefmark{1} (Senior Member, IEEE)}
\affil{Graduate School of Engineering, The University of Osaka, Osaka 565-0871, Japan}
\corresp{Corresponding author: Ryosuke Ikura (email: r.ikura@sip.comm.eng.osaka-u.ac.jp).}
\authornote{This work is supported in part by JSPS KAKENHI under Grant 23K26110 and 23K17461, and JST AdCORP under Grant JPMJKB2307.}

\begin{abstract}
This paper proposes a dynamic sensor scheduling method for sensor networks.
In sensor network applications, we often need multiple equally-informative node subsets that are activated sequentially to make a sensor network robust against concentrated battery consumption and sensor failures.
In addition, quality of these subsets changes dynamically and thus we must adapt those changes.
To find those node subsets, we propose a graph node partitioning method based on sampling theory for graph signals.
We aim to minimize the average reconstruction error for signals obtained at all node subsets, in contrast to conventional single subset selection. 
The graph node partitioning problem is formulated as a difference-of-convex (DC) optimization based on a subspace prior of graph signals, and is solved by the proximal DC algorithm. It guarantees convergence to a critical point.
To accommodate the online scenario where the signal subspace and optimal partitioning may change over time, we adaptively estimate the signal subspace from historical data and sequentially update the prior for our partitioning method.
Numerical experiments on synthetic and real-world sensor network data demonstrate that the proposed method achieves lower average mean squared errors compared to alternative methods.
\end{abstract}


\maketitle

\section{INTRODUCTION} 

\begin{table*}[t]
    \centering
    \caption{Comparison of graph node partitioning, sampling set selection, and clustering methods.}
    \label{tab:comparison}
    \begin{tabular}{l p{8.6cm} p{3cm}} 
    \toprule
        \textbf{Method} & \textbf{Primary Objective} & \textbf{Main Application} \\
    \midrule
        Graph Node Partitioning & 
        Dividing nodes into multiple equally-informative node subsets. & 
        Sensor scheduling\\ 
    \midrule
        Sampling Set Selection & 
        Selecting one subset of nodes that accurately reconstruct the whole signal. & 
        Efficient sensing\\
    \midrule
        Node Clustering & 
        Grouping nodes sharing similar properties. & 
        Network analysis \\ 
    \bottomrule
    \end{tabular}
\end{table*}

\IEEEPARstart{S}{ensor} networks have been used in various applications such as traffic, infrastructure, and facility monitoring systems~\cite{SNtraffic,SNinfra,SNfacility_monitor}. 
In practice, sensor networks often suffer from heavy power consumption and sensor failures during data collection and transmission~\cite{WSNenergy}.
To mitigate such risks, controlling sensor activations over time is crucial for many sensor network applications.

Sensor activation can be seen as a sensor selection problem at each time instance.
Its goal is to select a subset of $K$ sensors from $N$ candidates~($K < N$). 
The classical sensor placement problem often considers selecting static $K$ sensors.
However, this approach concentrates the sensing load on a fixed subset of nodes, which may shorten the lifespan of those sensors.
As an alternative, we can group sensors into disjoint subsets and activate them sequentially~\cite{Comparison}.
This strategy is known as \textit{sensor scheduling}~\cite{gupta2006stochastic}.
An effective sensor scheduling strategy must satisfy two essential requirements. 
\begin{enumerate}
    \item Accurate reconstruction: The whole signal can be accurately recovered from measurements obtained at active sensors at a given time. 
    \item Load balancing: Sensing loads are balanced among all sensors over time to avoid concentrated energy consumption.
\end{enumerate}
To satisfy these requirements, sensors must be partitioned into disjoint subsets, where each subset can accurately reconstruct the whole signal from its own measurements.
We consider this problem as graph node partitioning where sensor networks are mathematically represented as graphs.
Nodes and edges in a graph correspond to sensors and their connectivity, respectively.
Correspondingly, data collected through sensor networks can be modeled as \textit{graph signals}---discrete signals with their domain as nodes \cite{GSPsurvey1,GSPsurvey2,GSPsurvey3}.

Graph node partitioning relates to two other established approaches: sampling set selection and node clustering.
While all three approaches involve selecting or grouping node subsets, their objectives and applications differ fundamentally.
Table~\ref{tab:comparison} summarizes the key distinctions among them.

Sampling set selection of graph signals is widely studied in graph signal processing~\cite{isler2004sampling,sakiyamasan}.
In this approach, a designated number of nodes is selected so that the whole signal can be accurately reconstructed from the sampled measurements~\cite{SAMPsurvey}.
However, it only selects a single subset, and thus, for sensor scheduling, the activation load concentrates on the selected node subset.

Node clustering partitions nodes by assigning similar nodes to a group based on criteria such as cut minimization or submodularity maximization~\cite{spectral,SpectralSegmentation}.
While grouping similar nodes is beneficial for applications such as community detection, it is unsuitable for sensor scheduling.
This is because nodes in the same cluster may have similar measurements and thus it is challenging to reconstruct signals in a different cluster.

In summary, graph node partitioning is the strategy that satisfies the both requirements of high reconstruction accuracy and load balancing for sensor scheduling.

Existing graph node partitioning methods~\cite{Comparison, Tsitsvero} have the following limitations.
They typically rank nodes based on a predefined metric and sequentially assign them to subsets according to their ranking.
However, these strategies are heuristic and lack theoretical guarantees.
They also often rely on restrictive assumptions about the signal model, such as exact bandlimitedness.
Unless the assumptions are satisfied, their performance of reconstruction is not theoretically guaranteed.
More importantly, these methods focus exclusively on static graph node partitioning: They may not be suitable for direct application to \textit{dynamic sensor scheduling}, where signal statistics change over time.

In this paper, we propose a graph node partitioning method that overcomes the limitations of existing approaches.
The core idea is to reformulate graph node partitioning as a multiple sampling subsets selection~\cite{Comparison}, which naturally extends from a single sampling subset selection, to identify multiple disjoint node subsets, each capable of accurate signal reconstruction.
In addition, our formulation is free from the bandlimitedness assumption by using generalized sampling of graph signals based on a subspace prior~\cite{SAMPsurvey,tanaka2020generalized,harasanarbitrary}.

The single subset selection~\cite{SAMPsurvey} minimizes the reconstruction error for one subset.
We extend this to graph node partitioning by minimizing the average reconstruction error across all subsets.
This problem is encoded as a difference-of-convex (DC) optimization whose objective function is the difference of two or more convex functions~\cite{tao1998dc}.
This is solved by the proximal DC algorithm (PDCA) \cite{DC,PDCA}, which guarantees convergence to a critical point. 

To address online scenarios where the signal subspace is unknown and time-varying, we also propose a weighted dictionary learning scheme as an extension of the existing dictionary learning for graph signals~\cite{nomurasan}.
The existing method requires pre-training with the availability on the whole data to estimate an initial signal subspace. 
However, it is impractical for some applications.
In contrast, our extension enables robust subspace estimation without relying on the pre-training.

Experiments using both synthetic and real-world sensor network data demonstrate that the proposed method outperforms existing partitioning methods in terms of the mean squared error (MSE) of the reconstructed signals.

The remainder of this paper is organized as follows. 
Section~\ref{sec:related_work} reviews existing graph node partitioning methods that are related to this work.
In Section~\ref{sec:preliminary}, we introduce mathematical preliminaries for our method.
We propose our graph node partitioning method in Section~\ref{sec:proposed_method}.
Signal reconstruction experiments for synthetic and real-world signals
are presented in Section~\ref{sec:experiments}.
Section~\ref{sec:conclusion} concludes the paper.

\textit{Notation: } Bold lowercase and uppercase letters denote vectors and matrices, respectively.
We denote the $i$th column and $(i,j)$ element of a matrix $\mathbf{X}$ by $[\mathbf{X}]_{i}$ and $[\mathbf{X}]_{ij}$, respectively.
Similarly, $[\bm{x}]_i$ is an $i$th element of vector $\bm{x}$.
The operator $\operatorname{diag}(\mathbf{v})$ denotes the diagonal matrix with the elements of vector $\mathbf{v}$ on its diagonal, whereas $\operatorname{Diag}(\mathbf{V})$ extracts the diagonal elements of a square matrix $\mathbf{V}$ as a vector. The $\ell_2$-norm and Frobenius norm are denoted by $\|\cdot\|$ and $\|\cdot\|_F$, respectively. Calligraphic letters represent sets of indices; for a set $\mathcal{B}$, its complement is denoted by $\mathcal{B}^c$. The superscripts $^{\top}$ and $^{\dag}$ denote the transpose and Moore-Penrose pseudo-inverse, respectively.
The operator $\nabla$ represents the gradient.
In addition, $\odot$ represents the element-wise product.

\section{RELATED WORK}\label{sec:related_work}

In this section, we briefly review two existing graph node partitioning methods for sensor scheduling: Selection on relevance~(SRel) and that based on minimum Frobenius norm~(SFrob)~\cite{Comparison}.

Both SRel and SFrob share a common two-step strategy: (1) ranking all nodes based on a predefined importance criterion, and then (2) sequentially assigning these ranked nodes to partitioned subsets in a cyclic manner. The primary distinction lies in the definition of the importance of nodes. We describe them below.

\begin{description}
 
\item[\textbf{SRel:}] It ranks nodes based on the topology of the graph. This approach first performs node clustering by maximizing modularity~\cite{modularity1, modularity2}. Then, within each cluster, nodes are sorted based on the eigenvector centrality and sorted nodes are assigned to sampling subsets according to the rank. However, since this approach relies solely on structural features and disregards the characteristics of signals, its partitioning may be suboptimal.

\item[\textbf{SFrob:}] In contrast, it considers signal properties in the ranking algorithm. This approach extends the sampling theory for bandlimited graph signals~\cite{Tsitsvero} to graph node partitioning. Specifically, it ranks nodes according to their individual contribution to reducing the reconstruction error and assigns them to partitioned subsets following the rank. While this approach incorporates signal properties, its applicability is limited because of the bandlimited assumption.

\end{description}

In summary, existing approaches either ignore signal characteristics or rely on restrictive assumptions on signal subspaces.
Additionally, both are designed for static settings and do not adapt when signal characteristics change over time.
Thus, they are unsuitable for many real-world applications.

\section{PRELIMINARIES}\label{sec:preliminary}

In this section, we introduce a sampling framework for graph signals.
First, we present the graph signal sampling theory under a subspace prior.
Second, we describe the sampling set selection criterion based on the sampling framework .

We first introduce the graph basics.
We consider a weighted undirected graph $\mathcal{G}=(\mathcal{V},\mathcal{E})$, where $\mathcal{V}\ (|\mathcal{V}|=N)$ and $\mathcal{E}$ denote the set of nodes and edges, respectively. The adjacency matrix of $\mathcal{G}$ is denoted by $\bm{\mathbf{W}}$, where its $(i,j)$ element is the weight of the edge between the $i$th and $j$th nodes if they are connected, and 0 otherwise. The degree matrix $\mathbf{D}$ is defined as $\mathbf{D}=\text{diag}(d_0,d_1,\cdot \cdot \cdot,d_{N-1})$, where $d_n = \sum_m \mathbf{W}_{nm}$. We use graph Laplacian $\mathbf{L}=\mathbf{D}-\mathbf{W}$ as a graph variation operator~\cite{GSPsurvey3}. The graph signal $\bm{x} \in \mathbb{R}^N$ is defined as a mapping from the node set to the set of real numbers, i.e., $\bm{x}:\mathcal{V}\longrightarrow \mathbb{R}$.

The graph Fourier transform~(GFT) of $\bm{x}$ is defined as $\hat{\bm{x}}=\mathbf{U}^\top\bm{x}$ where the GFT matrix $\mathbf{U}$ is obtained by the eigen-decomposition of the graph Laplacian $\mathbf{L}=\mathbf{U}\mathbf{\Lambda}\mathbf{U}^\top$ with the eigenvalue matrix $\mathbf{\Lambda}=\text{diag}(\lambda_0,\lambda_1,\dots\lambda_{N-1})$. We refer to $\lambda_i$ as the \textit{$i$th graph frequency}.

\subsection{Graph Signal Sampling Theory Under Subspace Prior}

Subspace prior is a fundamental model in graph signal sampling theory, in which graph signals are assumed to lie in a known subspace \cite{SAMPsurvey}.
This includes the well-known bandlimited model as a special case.

A graph signal subspace is defined as follows \cite{SAMPsurvey}:
\begin{equation}
    \mathcal{A} \coloneqq \lbrace \bm{x}\mid \bm{x} = \mathbf{A}\bm{d} \text{ for }\bm{d} \in \mathbb{R}^M \rbrace,\label{eq:subspace_prior}
\end{equation}
where $\mathbf{A}\in\mathbb{R}^{N \times M}$, $M\le N$, is a generation transform depending on a graph, and $\mathbf{d}\in\mathbb{R}^{M}$ is a vector composed of expansion coefficients.

Here, we define a node domain sampling operator as follows:

\begin{definition}[Node domain sampling~\cite{SAMPsurvey}]
\textit{Let $\mathbf{I}_{\mathcal{MV}} \in \lbrace 0,1 \rbrace^{K \times N}$ be the submatrix of the identity matrix indexed by $\mathcal{M}\subset \mathcal{V}\ (|\mathcal{M}|=K)$ and $\mathcal{V}$. The sampling operator is defined as follows:}
\begin{equation}
    \mathbf{S}^{\top}\coloneqq\mathbf{I}_{\mathcal{MV}}\mathbf{G},\label{eq:samp_operator}
\end{equation}
\textit{where $\mathbf{G}\in\mathbb{R}^{N \times N}$ is an arbitrary linear graph filter. 
Thus, a sampled graph signal is given by $\bm{y} = \mathbf{S}^{\top} \bm{x}$.}
\end{definition}

In this paper, we consider the following noisy measurement model
\begin{equation}
    \bm{y} = \mathbf{S}^{\top} \bm{x} + \bm{\eta}, \label{sampling}
\end{equation}
where $\bm{\eta} \sim \mathcal{N}(\bm{0},\sigma^2\mathbf{I})$ is additive white Gaussian noise with standard deviation $\sigma$.

Under the subspace prior in \eqref{eq:subspace_prior}, the best possible recovery is obtained by solving the following minimax problem \cite{harasan}.
\begin{equation}
    \tilde{\bm{x}} = \underset{\tilde{\bm{x}} \in \mathcal{A}}{\operatorname{argmin}} \underset{\mathbf{S}^{\top} \bm{x}=\bm{y}}{\operatorname{max}}\| \tilde{\bm{x}} - \bm{x} \|^2 = \mathbf{A}(\mathbf{S}^{\top}\mathbf{A})^\dagger \bm{y}.\label{eq:mx_solution}
\end{equation}
Under the noiseless case, perfect recovery, i.e., $\bm{x}=\tilde{\bm{x}}$ is achieved when $\mathbf{S}^{\top}\mathbf{A}$ is invertible.
The condition is so-called \textit{direct sum condition} \cite{SAMPsurvey}.

\begin{figure*}[tbp]
    \centering
    \includegraphics[width=0.80\linewidth]{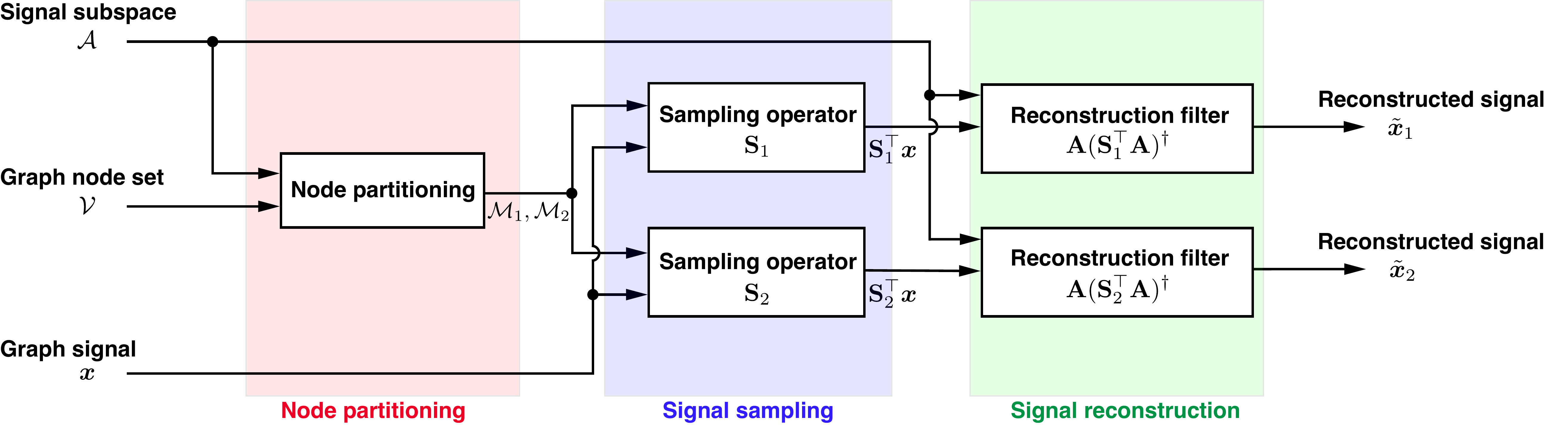}
    \caption{Overview of the proposed method. For simplicity, the selection with two subsets is illustrated.}\label{fig:buildingblock}
\end{figure*}

\subsection{Sampling Set Selection}
According to \eqref{eq:mx_solution}, the expected value of MSE is upper bounded by the following relationship~\cite{harasan}:
\begin{equation}
\begin{split} \label{eq:mx_solution_bound}
&\mathbb{E}\lbrack \| \tilde{\bm{x}} - \bm{x} \|^2 \rbrack  \\
&=\text{tr}(\bm{x}\bm{x}^{\top} - \mathbf{E}\bm{x}\bm{x}^{\top}\mathbf{E}^{\top}) + \text{tr}(\mathbf{A}(\mathbf{S}^{\top}\mathbf{A})^\dagger \mathbf{\Gamma}_{\eta} (\mathbf{A}^{\top}\mathbf{S})^\dagger \mathbf{A}^{\top}) \\
&\leq\text{tr}(\mathbf{\Gamma}_{\eta})\text{tr}(\mathbf{A}^{\top}\mathbf{A})\text{tr}((\mathbf{S}^{\top}\mathbf{A}\mathbf{A}^{\top}\mathbf{S})^{-1}),
\end{split}
\end{equation}
where $\mathbf{E} = \mathbf{A}(\mathbf{S}^{\top}\mathbf{A})^\dagger \mathbf{S}^{\top}$ and $\mathbf{\Gamma}_{\eta} = \mathbb{E}[\bm{\eta\eta}^{\top}]$.
Hereafter, we suppose that $\mathbf{S}^{\top}\mathbf{A}\mathbf{A}^{\top}\mathbf{S}$ is invertible for simplicity\footnote{The same formulation can be easily derived even if not invertible.}.

To minimize \eqref{eq:mx_solution_bound}, the optimal sampling set $\mathcal{M}$ can be obtained by solving the following problem.

\begin{equation}
\begin{split}
\mathcal{M}^*&=\underset{\mathcal{M}\subset\mathcal{V}}{\operatorname{argmin}}~\text{tr}
((\mathbf{S}^\top\mathbf{A}\mathbf{A}^\top\mathbf{S})^{-1}). 
\label{A-optimality}
\end{split}
\end{equation}
This problem is a combinatorial optimization and NP-hard.
Therefore, existing sampling set selection methods typically perform a greedy selection~\cite{AOptimalityGreedy1,AOptimalityGreedy2}, which yields suboptimal solutions in general.

Note that even if $\mathcal{M}^*$ in \eqref{A-optimality} is the global optimum, the remaining subset $\mathcal{M}^c = \mathcal{V}\backslash \mathcal{M}^*$ is generally not an equally-informative subset compared to $\mathcal{M}^*$.
Therefore, sampling set selection cannot be applied to graph node partitioning straightforwardly.

\section{GRAPH NODE PARTITIONING AND ONLINE SENSOR SCHEDULING} \label{sec:proposed_method} %

In this section, we first present the static version of the proposed graph node partitioning based on graph signal sampling theory. Next, we extend it to the online sensor scheduling problem where the optimal partitioning can vary over time.
Finally, we formulate a dictionary learning problem for estimating a time-varying signal subspace from the observed data.

\subsection{Static Graph Node Partitioning}
Fig.~\ref{fig:buildingblock} illustrates the overview of the proposed static graph node partitioning.
For simplicity, we describe the case of bipartitioning with noiseless observations.
It consists of three stages: Graph node partitioning, signal sampling, and signal reconstruction. First, with the given signal subspace $\mathcal{A}$ in~\eqref{eq:subspace_prior}, the node set $\mathcal{V}$ is divided into subsets $\mathcal{M}_k$. Based on this partitioning, the sampling operators $\mathbf{S}_k$ are determined to sample the graph signal $\bm{x}$. Finally, the full graph signal is reconstructed from the sampled measurements $\mathbf{S}_k^\top \bm{x}$ in~\eqref{sampling} using the reconstruction filter $\mathbf{A}(\mathbf{S}_k^\top \mathbf{A})^\dagger$ in~\eqref{eq:mx_solution}.

Our primary contribution is the design of efficient graph node partitioning for both static and time-varying signal subspaces.
Accordingly, we employ existing methods~\cite{SAMPsurvey,tanaka2020generalized} for the signal sampling and reconstruction.
We describe the graph node partitioning in the following.

\subsubsection{Problem Formulation}
Here, we assume that the signal subspace $\mathcal{A}$ is given and it is specified by generation transform $\mathbf{A}$.
We focus on a bipartitioning scenario $\mathcal{M}_1,\mathcal{M}_2\subset\mathcal{V}$ where all subsets are non-overlapping and the number of nodes in each subset is equal, i.e., $|\mathcal{M}_1|=|\mathcal{M}_2|=N/2$.\footnote{For odd $N$, one subset has $\lceil N/2\rceil$ nodes and the other one has $\lfloor N/2\rfloor$ nodes.}
Note that, it can be applied to the $2^k$ partitioning by hierarchically cascading the bipartitioning to the resulting subsets.

The sampling operators for $\mathcal{M}_1$ and $\mathcal{M}_2$ are defined as $\mathbf{S}^\top_1=\mathbf{I}_{\mathcal{M}_1\mathcal{V}}$ and $\mathbf{S}^\top_2=\mathbf{I}_{\mathcal{M}_2\mathcal{V}}$,
respectively, where we set $\mathbf{G}=\mathbf{I}$ in \eqref{eq:samp_operator} for simplicity.
According to \eqref{sampling}, two sampled signals are expressed as $\bm{y}_1 = \mathbf{S}_1^\top\bm{x} + \bm{\eta}_1$ and $\bm{y}_2 = \mathbf{S}_2^\top\bm{x}+ \bm{\eta}_2$.
Similar to \eqref{eq:mx_solution}, minimax recovery solutions of each observation are obtained as $\tilde{\bm{x}}_1=\mathbf{A}(\mathbf{S}^\top_1\mathbf{A})^\dagger\bm{y}_1$ and $\tilde{\bm{x}}_2=\mathbf{A}(\mathbf{S}^\top_2\mathbf{A})^\dagger\bm{y}_2$.

Our purpose is to minimize the \textit{average} reconstruction errors
across all subsets.
Based on \eqref{eq:mx_solution_bound}, the average reconstruction error is upper bounded as
\begin{equation}
\begin{split} \label{eq:mx_solution_bound_average}
&\frac{1}{2}\left(\mathbb{E}\lbrack \| \tilde{\bm{x}}_1 - \bm{x} \|^2+\mathbb{E}\lbrack \| \tilde{\bm{x}}_2 - \bm{x} \|^2 \rbrack\right)  \\
&\leq C\left(\text{tr}((\mathbf{S}^{\top}_1\mathbf{A}\mathbf{A}^{\top}\mathbf{S}_1)^{-1})+\text{tr}((\mathbf{S}^{\top}_2\mathbf{A}\mathbf{A}^{\top}\mathbf{S}_2)^{-1})\right),
\end{split}
\end{equation}
where $C=\frac{1}{2}\text{tr}(\mathbf{\Gamma}_{\eta})\text{tr}(\mathbf{A}^{\top}\mathbf{A})$ is a constant.

Based on \eqref{A-optimality} and \eqref{eq:mx_solution_bound_average}, the following optimization problem can be considered for graph node partitioning:
\begin{equation}
\begin{gathered}
\begin{aligned}\label{eq:bipartition_set}
&(\mathcal{M}^{\ast}_1, \mathcal{M}_2^{\ast}) = \\
&\underset{\mathcal{M}_1,\mathcal{M}_2 \subset \mathcal{V}} {\operatorname{argmin}}\ \text{tr}((\mathbf{S}_1^{\top}\mathbf{A}\mathbf{A}^{\top}\mathbf{S}_1)^{-1}) +\text{tr}((\mathbf{S}_2^{\top}\mathbf{A}\mathbf{A}^{\top}\mathbf{S}_2)^{-1})\\
\end{aligned} \\
\text{ s.t.~} \mathcal{M}_1\cap\mathcal{M}_2=\emptyset, \, |\mathcal{M}_1|=|\mathcal{M}_2|=\frac{N}{2}.
\end{gathered}
\end{equation}
Similar to the single subset selection, it is combinatorial and NP-hard.
Furthermore, it involves matrix inversion that requires huge computational burden.

For tractability, we first approximate the calculation of the matrix inverses with the second-order Neumann series approximation~\cite{Neumann}.
The details are shown in Appendix. %
As a result, \eqref{eq:bipartition_set} is approximated as 
\begin{equation}
\begin{gathered} \label{bipartition2}
\begin{aligned}
    &(\mathcal{M}^{\ast}_1,\mathcal{M}_2^{\ast}) = \\
    &\underset{\mathcal{M}_1,\mathcal{M}_2 \subset \mathcal{V}} {\operatorname{argmin}}\text{tr}((\mathbf{S}_1^\top\mathbf{AA}^\top\mathbf{S}_1)^2 + (\mathbf{S}_2^\top\mathbf{AA}^\top\mathbf{S}_2)^2)
    \end{aligned} \\
\text{s.t.~} \mathcal{M}_1\cap\mathcal{M}_2=\emptyset, \, |\mathcal{M}_1|=|\mathcal{M}_2|=\frac{N}{2}.
\end{gathered}
\end{equation}

Let $\bm{m}_k\in\{0,1\}^{N}$ ($k \in \{1,2\}$) be the indicator vectors, whose $i$th element $[\bm{m}_k]_i$ is defined as
\begin{equation}
    [\bm{m}_k]_i = \begin{cases}
        1 & \text{if } i\in\mathcal{M}_k,\\
        0 & \text{otherwise.}
    \end{cases} \label{m_sampling_operator}
\end{equation}
Then, we further rewrite \eqref{bipartition2} with the cyclic property of trace and $\bm{m}_k$ as follows:
\begin{equation}
\begin{gathered}
\begin{aligned}\label{eq:bipartition_binary_2}
    &\bm{m}^{\ast}_1=\\ 
    &\underset{\bm{m}_1\in \{0,1\}^N} {\operatorname{argmin}}\ \text{tr}((\mathbf{A}^{\top} \text{diag}(\bm{m}_1)\mathbf{A})^2) + \text{tr}((\mathbf{A}^{\top} \text{diag}(\bm{1}-\bm{m}_1)\mathbf{A})^2)
\end{aligned}\\
    \text{s.t.~}\bm{m}_1^\top(\bm{1}-\bm{m}_1)=0, \, \bm{1}^\top\bm{m}_1 = \frac{N}{2},
\end{gathered}    
\end{equation}
where we use the relationship $\bm{m}_1 + \bm{m}_2 = \bm{1}$.

Note that \eqref{eq:bipartition_binary_2} is still combinatorial due to the binary $\bm{m}$. 
Therefore, we introduce a convex relaxation of \eqref{eq:bipartition_binary_2} by considering a continuous $\bm{m}_{\text{relaxed}}\in [0,1]^N$ instead of $\bm{m}$.
Finally, the problem to be solved is represented as follows.
\begin{equation}
\begin{gathered}
\begin{split}\label{eq:bipartition_continuous}
    \bm{m}_{\text{relaxed}}^{\ast} 
    =\underset{\bm{m}_{\text{relaxed}}\in [0,1]^N} {\operatorname{argmin}}\ &\text{tr}((\mathbf{A}^{\top} \text{diag}(\bm{m}_{\text{relaxed}})\mathbf{A})^2) \\
     &+\text{tr}(( \mathbf{A}^{\top} \text{diag}(\bm{1}-\bm{m}_{\text{relaxed}})\mathbf{A})^2)
\end{split}\\    
    \text{s.t. }\bm{m}_{\text{relaxed}}^\top(\bm{1}-\bm{m}_{\text{relaxed}})=0, \, \bm{1}^\top\bm{m}_{\text{relaxed}}=\frac{N}{2}.
\end{gathered}
\end{equation}
Although the relaxed variables reside in the convex set $[0,1]^N$, \eqref{eq:bipartition_continuous} is non-convex due to the constraint $\bm{m}_{\text{relaxed}}^\top(\bm{1}-\bm{m}_{\text{relaxed}})=0$.
However, as this constraint function is a DC function, the solution presented below leads to a critical point.

\subsubsection{Solver}

We reformulate \eqref{eq:bipartition_continuous} into the applicable form to PDCA \cite{PDCA} as follows:
\begin{equation}
   \bm{m}_{\text{relaxed}}^{\ast}=\underset{\bm{m}_{\text{relaxed}}}{\operatorname{argmin}}
    ~f(\bm{m}_{\text{relaxed}})+g(\bm{m}_{\text{relaxed}})-h(\bm{m}_{\text{relaxed}}),\label{DC_form}
\end{equation}
where $f(\bm{m}_{\text{relaxed}})$ and $h(\bm{m}_{\text{relaxed}})$ are differentiable convex, and $g(\bm{m}_{\text{relaxed}})$ is non-differentiable but convex.
In addition, $g$ is proximable, i.e., whose proximity operator, which is defined as
\begin{equation} \label{def_proximity_operator}
    \text{prox}_{\gamma g}(\bm{m}_{\text{relaxed}}) \coloneqq \underset{\bm{y}}{\operatorname{argmin}}~g(\bm{y}) + \frac{1}{2\gamma}\|\bm{m}_{\text{relaxed}}-\bm{y}\|_2^2,
\end{equation}
can be solved efficiently with high precision~\cite{condat2013primal}.

We define functions in~\eqref{DC_form} as follows.
\begin{equation}
\begin{split}
    &f(\bm{m}_{\text{relaxed}}) = \text{tr}((\mathbf{A}^\top \text{diag}(\bm{m}_{\text{relaxed}}) \mathbf{A})^2) \\
    &\hspace{5.7em}+\text{tr}((\mathbf{A}^\top \text{diag}(\bm{1}-\bm{m}_{\text{relaxed}}) \mathbf{A})^2),\\
    &g(\bm{m}_{\text{relaxed}})=  \iota_{\mathcal{C}_{\text{card}}}(\bm{m}_{\text{relaxed}})+ \iota_{\mathcal{C}_\text{box}}(\bm{m}_{\text{relaxed}}), \\
    &h(\bm{m}_{\text{relaxed}}) = \beta(\bm{1}^\top(\bm{m}_{\text{relaxed}}\odot\bm{m}_{\text{relaxed}})-\bm{1}^\top \bm{m}_{\text{relaxed}}), \label{convex_functions}
\end{split}
\end{equation}
where $\beta$ is the parameter.
In addition, the indicator function is defined as
\begin{equation}
    \iota_\mathcal{C}(\bm{m}_{\text{relaxed}}) = 
    \begin{cases}
        0 & \text{if } \bm{m}_{\text{relaxed}}\in \mathcal{C} \\
        +\infty & \text{otherwise},
    \end{cases}\label{eq:cardi_const}
\end{equation}
where $\mathcal{C}$ is a convex set.
We can convert the hard constraint $\bm{1}^\top\bm{m}_{\text{relaxed}}=\frac{N}{2}$ in \eqref{eq:bipartition_continuous} into the objective function in \eqref{convex_functions} by $\mathcal{C}_{\text{card}}=\{\bm{m}_{\text{relaxed}}\mid\bm{1}^\top\bm{m}_{\text{relaxed}}=\frac{N}{2}\}$ with~\eqref{eq:cardi_const}.
We also constrain $\bm{m}_{\text{relaxed}}\in[0,1]^N$ as $\mathcal{C}_{\text{box}}=\{ \bm{m}_{\text{relaxed}}\mid\bm{m}_{\text{relaxed}}\in[0,1]^N\}$.
With an appropriate choice of $\beta$, \eqref{DC_form} becomes identical to \eqref{eq:bipartition_continuous} \cite{rockafellar1997convex}.

Algorithm~\ref{alg1} shows the detailed steps for solving \eqref{DC_form}. We use the following operators in the algorithm.

\begin{equation}
\begin{split}
    \nabla f(\bm{m}_{\text{relaxed}}) &=2\text{Diag}(\mathbf{AA}^\top(2\text{diag}(\bm{m}_{\text{relaxed}})-\mathbf{I})\mathbf{A}\mathbf{A}^\top), \\
    \nabla h(\bm{m}_{\text{relaxed}}) &= \beta(2\bm{m}_{\text{relaxed}} - \bm{1}).
\end{split}
\end{equation}
The computation of $\text{prox}_{\gamma g}$ in~\eqref{def_proximity_operator} is a convex optimization since $g(\bm{m}_{\text{relaxed}})$ consists of multiple convex functions.
Therefore, it can be solved via an existing convex solver. Specifically, we use alternating direction method of multipliers~(ADMM)~\cite{ADMM}.
Since the resulting $\bm{m}_{\text{relaxed}} \in [0,1]^N$ is a real-valued vector, we binarize it by thresholding at the end of the algorithm to obtain the binary vector $\bm{m}\in \lbrace0,1\rbrace^N$.

\begin{algorithm}[tbp]
	\caption{Static graph node partitioning}
    \label{alg1}
    \begin{algorithmic}[1]
    \REQUIRE $\bm{m}_{\text{relaxed}}^{(0)} \in [0,1]^N$, Lipschitz constant $L > 0$, signal subspace $\mathbf{A}$.
    \STATE Set step size $\gamma \leftarrow 1/L$.
    \STATE $k \leftarrow 0$
    \WHILE{convergence criterion is not met}
        \STATE Compute a gradient $\bm{u}^{(k)} \leftarrow \nabla h(\bm{m}_{\text{relaxed}}^{(k)})$.
        \STATE Update variable:
        \begin{equation*}
            \bm{m}_{\text{relaxed}}^{(k+1)} \leftarrow \text{prox}_{\gamma g} ( \bm{m}_{\text{relaxed}}^{(k)} - \gamma ( \nabla f(\bm{m}_{\text{relaxed}}^{(k)}) - \bm{u}^{(k)}))
        \end{equation*}
        \STATE $k \leftarrow k+1$
    \ENDWHILE
    \STATE Binarize the continuous vector by thresholding:
    \begin{equation}
        [\bm{m}]_i = 
        \begin{cases}
            1\quad\text{if }[\bm{m}_{\text{relaxed}}^{(k+1)}]_i >\frac{1}{2}\\
            0\quad\text{otherwise.}
        \end{cases}
    \end{equation}
    \STATE \textbf{Output: } $\bm{m}\in\lbrace0,1\rbrace^N$
    \end{algorithmic}
\end{algorithm}

\subsection{Online Graph Node Partitioning}
We extend the static graph node partitioning to the online scenario. Since the signal subspace may be time-varying in this setting, the optimal graph node partitioning also varies at each time instance.

We use the subscript $t$ to specify the time instance of variables.
For example, the signal subspace at $t$ is defined by $\mathbf{A}_t$.

We describe the case of $M$ partitioning $(M=2^k, k=1,2,\ldots)$ in the following.
Here, we assume that $\lbrace \mathbf{A}_t \rbrace$ is given.
In our online graph node partitioning, the following process is performed at every $M$ time instances.

First, $\mathcal{V}$ is partitioned into $M$ disjoint subsets based on the current signal subspace $\mathbf{A}_t$.
Specifically, by recursively executing the Algorithm~\ref{alg1}, we obtain partitioned node subsets $\lbrace\mathcal{M}_i\rbrace_{i=1}^M$ satisfying $\mathcal{M}_1 \oplus \ldots \oplus\mathcal{M}_M = \mathcal{V}$.

Second, during the subsequent $M$ time instances, we perform signal sampling on the nodes associated with the specific subset selected at each time step, i.e.,
\begin{equation} \label{sequential_sampling}
    \bm{y}_t  = \mathbf{S}_t^\top \bm{x}_t + \bm{\eta}_t.
\end{equation}
By defining the index~$l=((t-1)~(\text{mod }M)+1)$, the sampling operator $\mathbf{S}_t$ corresponds to the subset $\mathcal{M}_l$.
In addition, $\bm{x}_t$ represents the original signal at $t$.

Finally, as shown in~\eqref{eq:mx_solution}, the whole signal is reconstructed using the corresponding signal subspaces $\mathbf{A}_t$, i.e.,

\begin{equation} \label{sequential_signal_reconstruction}
\begin{split}
    \tilde{\bm{x}}_{t}=\mathbf{A}_t(\mathbf{S}_t^\top \mathbf{A}_t)^\dagger \bm{y}_t.
\end{split}
\end{equation}

\subsection{Online Dictionary Learning} \label{subsec:online_dictionary_learning}
In the previous subsections, the signal subspace $\mathbf{A}$ or $\lbrace \mathbf{A}_t \rbrace$ is assumed to be given. 
However, it is not explicitly provided in general.
Therefore, we would like to adaptively estimate $\lbrace \mathbf{A}_t \rbrace$ from the previously reconstructed signals from $t-D$ to $t-1$, which is written as $\tilde{\mathbf{X}}_{t-1} = [\tilde{\bm{x}}_{t-D},\ldots,\tilde{\bm{x}}_{t-1}]$.
This problem can be referred to as subspace tracking.

Since $\tilde{\bm{x}}_{\tau}$ in~\eqref{sequential_signal_reconstruction} contains both measurements at reliable nodes and potentially inaccurate reconstructed signals, we propose to differentiate between them by introducing a confidence matrix that weights the data fidelity term based on the sampling history.

Let $\mathbf{D}_{t-1} = [\bm{d}_{t-D},\dots,\bm{d}_{t-1}] \in \mathbb{R}^{N \times D}$ be the collection of estimated expansion coefficients.
Under the subspace prior in~\eqref{eq:subspace_prior}, we can express $\tilde{\mathbf{X}}_{t-1} \approx \mathbf{A}_{t-1} \mathbf{D}_{t-1}$.
We formulate the following subspace tracking problem as a dictionary learning problem:
\begin{equation}
\begin{split}
    (\mathbf{A}_t^*, \mathbf{D}_t^*) = \underset{\mathbf{A}_t,\mathbf{D}_t}{\operatorname{argmin}}&\sum_{i=0}^{D-1}\|\mathbf{W}_i([\tilde{\mathbf{X}}_{t-1}]_i-\mathbf{A}_t[\mathbf{D}_t]_i)\|_F^2 \\
    &\text{s.t. }\|\mathbf{D}_t\|_1 \leq K, \label{dictionary_learning}
\end{split}
\end{equation}
where the confidence matrix $\mathbf{W}_i$ is defined as
\begin{equation}
    \mathbf{W}_i = \text{diag}(\bm{w}_i).
\end{equation}
The vector $\bm{w}_i$ denotes a confidence vector whose $i$th element is the confidence weight of the $i$th node.
It could be determined by the sampling pattern at the corresponding time instance, such as high confidence at sampled signals and low confidence at unsampled ones.
In~\eqref{dictionary_learning}, we impose a sparsity constraint on $\mathbf{D}_t$ which is a similar setting to the well-studied dictionary learning problems~\cite{nomurasan,dictionary_learning_1,dictionary_learning_2}.

The fundamental distinction between the proposed formulation and the existing online dictionary learning method for graph signals~\cite{nomurasan} lies in the introduction of the weighting matrix $\mathbf{W}_i$.
The method in~\cite{nomurasan} assumes that $\mathbf{A}_t$ changes smoothly over time and includes the regularization term $\|\mathbf{A}_t - \mathbf{A}_{t-1}\|_F^2$ to ensure a stable learning.
In contrast, our formulation utilizes $\mathbf{W}_i$ to ensure that the subspace is learned primarily from measurements at reliable nodes.
This mechanism prevents error propagation from inaccurate reconstructions and stabilizes the learning process, thereby allowing us to omit the temporal regularization term.

We decompose the optimization problem in~\eqref{dictionary_learning} into two independent subproblems with respect to $\mathbf{A}_t$ and $\mathbf{D}_t$, and solve them alternately, similar to the approach described in~\cite{nomurasan}.

First, we optimize \eqref{dictionary_learning} with respect to $\mathbf{D}_t$ by fixing $\mathbf{A}_t$. 
The problem is formulated as:
\begin{equation} \label{eq:D_update_step}
    \mathbf{D}_t^* = \underset{\mathbf{D}_t}{\operatorname{argmin}}~\psi(\mathbf{D}_t) + \iota_{\mathcal{C}_{\text{sparse}}}(\mathbf{D}_t),
\end{equation}
where $\psi(\mathbf{D}_t) = \sum_{i=0}^{D-1}\|\mathbf{W}_i([\tilde{\mathbf{X}}_{t-1}]_i-\mathbf{A}_t[\mathbf{D}_t]_i)\|_F^2$ and $\iota_{\mathcal{C}_{\text{sparse}}}$ is an indicator function in~\eqref{eq:cardi_const}.
Here, $\mathcal{C}_{\text{sparse}}$ is defined as:
\begin{equation}
    \mathcal{C}_{\text{sparse}} = \lbrace \mathbf{D}~|~\|\mathbf{D}\|_1 \leq K \rbrace. \label{sparsity_constraint}
\end{equation}
Note that~\eqref{eq:D_update_step} is a convex optimization where the first term is differentiable and the second term is proximable.
We use proximal gradient descent~\cite{proximal_gradient_method} to solve it.
The optimal solution is obtained by iteratively performing the following update until convergence:
\begin{equation} 
    \mathbf{D}_t^{n+1} = \operatorname{prox}_{\iota_{\mathcal{C}_{\text{sparse}}}} (\mathbf{D}_t^n - \gamma \nabla \psi(\mathbf{D}_t^n)),
\label{D_update}
\end{equation}
where the superscript $n+1$ is the iteration number and $\gamma$ is the step size.
The gradient $\nabla \psi(\mathbf{D}_t)$ and the proximity operator $\operatorname{prox}_{\iota_{\mathcal{C}_{\text{sparse}}}}$ are computed as follows.
We calculate $\nabla\psi(\mathbf{D}_t^n)$ column-by-column.
For the $i$th column ($i=0,\dots,D-1$), the gradient is given by
\begin{equation}
    \nabla_{[\mathbf{D}_t]_i}\psi(\mathbf{D}_t) = -2\mathbf{A}_t^\top \mathbf{W}_i^\top \mathbf{W}_i([\tilde{\mathbf{X}}_{t-1}]_i-\mathbf{A}_t[\mathbf{D}_t]_i).
\end{equation}
Followed by Moreau's decomposition~\cite{combettes2013moreau}, the proximity operator is computed via that of $\ell_\infty$-norm, i.e., 
\begin{equation} \label{D_prox}
\text{prox}_{\iota_{\mathcal{C}_{\text{sparse}}}}(\mathbf{Y}) = \mathbf{Y}-\text{prox}_{K\|\cdot\|_\infty}(\mathbf{Y}).
\end{equation}
In addition, the second term in~\eqref{D_prox} is computed element-wise as
\begin{equation}
    [\text{prox}_{K\|\cdot\|_\infty}(\mathbf{Y})]_{ij} = \text{sign}([\mathbf{Y}]_{ij}) \min\left\{ |[\mathbf{Y}]_{ij}|, \zeta_i \right\},
\end{equation}
where $\zeta_i$ is the unique solution to the following equation
\begin{equation}
    \sum_{j=0}^{D-1}\max\left\{ 0, |[\mathbf{Y}]_{ij}|-\zeta_i \right\} = \frac{K}{N}.
\end{equation}

Second, we update $\mathbf{A}_t$ by solving~\eqref{dictionary_learning} with fixed $\mathbf{D}_t$, i.e., 
\begin{equation}
    \mathbf{A}_t^* =  \underset{\mathbf{A}_t}{\operatorname{argmin}}\sum_{i=0}^{D-1}\|\mathbf{W}_i([\tilde{\mathbf{X}}_{t-1}]_i-\mathbf{A}_t [\mathbf{D}_t]_i)\|_F^2. \label{A_step}
\end{equation}
We solve~\eqref{A_step} by using gradient descent method~\cite{gradient_descent}. The gradient of objective function in~\eqref{A_step} is given by
\begin{equation}
\begin{split}
        &\nabla_{\mathbf{A}_t}\sum_{i=0}^{D-1}\|\mathbf{W}_i ([\tilde{\mathbf{X}}_{t-1}]_i-\mathbf{A}_t[\mathbf{D}_t]_i)\|_F^2 \\
        &\quad = 2\sum_{i=0}^{D-1}\mathbf{W}_i^\top \mathbf{W}_i(\mathbf{A}_t [\mathbf{D}_t]_i-[\tilde{\mathbf{X}}_{t-1}]_i [\mathbf{D}_t]_i^\top).
\end{split}
\end{equation}

We repeat these two steps until convergence. Algorithm~\ref{alg3} summarizes its algorithm.

\begin{algorithm}[tbp]
\caption{Dictionary learning with confidence matrix}
\label{alg3}
\begin{algorithmic}[1]
\REQUIRE $\tilde{\mathbf{X}}_{t-1}$, $\{\mathbf{W}_i\}_{i=0}^{D-1}$, $K$, $\gamma_{D}, \gamma_{A}$
\STATE Initialize $\mathbf{A}_t \leftarrow \mathbf{A}_{t-1}$ and $\mathbf{D}_t \leftarrow \bm{1}\bm{1}^\top$.
\WHILE{convergence criterion is not met}
    \STATE Step 1: Update coefficients $\mathbf{D}_t$
    \STATE $n \leftarrow 0$, $\mathbf{D}_t^{(n)} \leftarrow \mathbf{D}_t$.
    \WHILE{convergence criterion for $\mathbf{D}$ not met}
        \STATE $ \mathbf{D}_t^{(n+1)} \leftarrow \operatorname{prox}_{\gamma_{D} \iota_{\mathcal{C}_{\text{sparse}}}} \left( \mathbf{D}_t^{(n)} - \gamma_{D} \nabla \psi(\mathbf{D}_t^{(n)}) \right)$
        \STATE $n \leftarrow n + 1$
    \ENDWHILE
    \STATE $\mathbf{D}_t \leftarrow \mathbf{D}_t^{(n)}$.

    \STATE Step 2: Update Dictionary $\mathbf{A}_t$
    \STATE $m \leftarrow 0$, $\mathbf{A}_t^{(m)} \leftarrow \mathbf{A}_t$.
    \WHILE{convergence criterion for $\mathbf{A}$ not met}
        \STATE  $\mathbf{G}_{A} \leftarrow 2\sum\limits_{i=0}^{D-1}\mathbf{W}_i^\top \mathbf{W}_i(\mathbf{A}_t^{(m)} [\mathbf{D}_t]_i-[\tilde{\mathbf{X}}_{t-1}]_i [\mathbf{D}_t]_i^\top)$
        \STATE $ \mathbf{A}_t^{(m+1)} \leftarrow \mathbf{A}_t^{(m)} - \gamma_{A}\mathbf{G}_{A}$
        \STATE $m \leftarrow m + 1$
    \ENDWHILE
    \STATE $\mathbf{A}_t \leftarrow \mathbf{A}_t^{(m)}$.
\ENDWHILE
\STATE \textbf{Output:} $\mathbf{A}_t$
\end{algorithmic}
\end{algorithm}

We can simply integrate the signal subspace learning steps with the proposed graph node partitioning. 
After a signal reconstruction in~\eqref{sequential_signal_reconstruction}, the reconstructed signal $\tilde{\bm{x}}_t$ is appended to the buffer of the reconstructed signals, i.e.,
\begin{equation} \label{buffer_update}
\tilde{\mathbf{X}}_{t}
=\begin{cases}    
[\tilde{\mathbf{X}}_{t-1}, \tilde{\bm{x}}_{t}]~&\text{ if}~t\leq D,\\
[[\tilde{\mathbf{X}}_{t-1}]_2,\dots,[\tilde{\mathbf{X}}_{t-1}]_D, \tilde{\bm{x}}_{t}]~&\text{ otherwise.}
\end{cases}
\end{equation}
Then, we perform the dictionary learning \eqref{dictionary_learning} to estimate the signal subspace for the next time instance.
The overall algorithm of the online sensor scheduling is summarized as Algorithm~\ref{alg2}.

In our implementation, we set the initial signal subspace as $\mathbf{A}_0 = \mathbf{I}$.
On the other hand, many alternative methods require it to be initialized more carefully~\cite{nomurasan,dictionary_learning_1,dictionary_learning_2}.
Typically, $\mathbf{A}_0$ is determined via the singular value decomposition of a given data.
This difference stems from the fact that~\eqref{dictionary_learning} can learn subspace stably over time with any choice of $\mathbf{A}_0$, which is beneficial for online sensor scheduling under insufficient initial observations~(i.e., cold start).

\begin{algorithm}[tbp]
    \caption{Online sensor scheduling}
    \label{alg2}
    \begin{algorithmic}[1]
    \REQUIRE  $K>0$, $M=2^k,~(k=1,2,\ldots)$, $\mathbf{A}_0 = \mathbf{I}$
    \STATE $t \leftarrow 0$.
    \WHILE{the sensor system is activated}
        \STATE \textit{Step 1: Graph Node Partitioning}
        \STATE Compute graph node partitioning $\lbrace \mathcal{M}_k \rbrace_{k=1}^M$ by solving \eqref{eq:bipartition_continuous}.
        \STATE \textit{Step 2: Sequential Sampling \& Reconstruction}
        \FOR{$k = 1, \ldots ,M$}
            \STATE Acquire graph signals from the subset $\mathcal{M}_{k}$.
            \STATE Reconstruct the full signal $\tilde{\bm{x}}_{t+k}$ via \eqref{sequential_signal_reconstruction}.
            \STATE Update the historical data buffer to obtain $\tilde{\mathbf{X}}_{t+k-1}$ according to \eqref{buffer_update}.
            \STATE Perform dictionary learning~\eqref{dictionary_learning} using $\tilde{\mathbf{X}}_{t+k-1}$ and update the signal subspace $\mathbf{A}_{t+k}$.
            \STATE $t \leftarrow t + 1$
        \ENDFOR
    \ENDWHILE
    \end{algorithmic}
\end{algorithm}

\section{EXPERIMENTS} \label{sec:experiments}

We validate the effectiveness of the proposed method via reconstruction experiments for synthetic and real-world graph signals. We conduct three experiments: Static partitioning on synthetic graph signals, online partitioning on synthetic graph signals, and online partitioning on real-world data.

\subsection{Static Partitioning on Synthetic Graph Signals} \label{experiment1}
First, in order to validate the effectiveness of our static graph node partitioning, we compare its performance to that of alternative graph node partitioning methods.
For this experiment, we use Algorithm~\ref{alg1}.
 \subsubsection{Graph and Signal Synthesis}

We generate a random sensor graph using the following procedure:
First, $256$ nodes are randomly distributed in the 2-D space $[0,1]\times[0,1]$.
For each node, we connect edges to its $k$ nearest neighbors and assign edge weights $\exp(-d^2)$ where $d$ is the Euclidean distance between two nodes.
To simulate realistic sensor networks~\cite{xue2004number}, $k$ is randomly chosen between two to eight for each node. 

For graph signals, we consider two full-band signals:
\begin{enumerate}
    \item \textbf{Heat-diffusion (HD) graph signal: }
    Its spectrum decays slowly as the graph frequency $\lambda$ increases. According to~\eqref{eq:subspace_prior}, this signal is expressed as
    \begin{equation} \label{BandLimitedSignal}
        \bm{x}_{\text{heat}} = \mathbf{A}_{\text{heat}}\bm{d},
    \end{equation}
    where $\mathbf{A}_{\text{heat}} = \mathbf{U}\widehat{H}(\mathbf{\Lambda})\mathbf{U}^\top$ with $\widehat{H}(\mathbf{\Lambda})=\text{exp}(-\alpha \mathbf{\Lambda})$. We set $\alpha = 10$ and $\bm{d}\sim\mathcal{N}(\bm{1},\mathbf{I})$.
    
    \item \textbf{Piecewise smooth (PWS) graph signal: }
    It forms clustered signals such that the signal has different average values in different clusters and also has smooth variations within the cluster.
    The PWS signal is therefore represented as:
    \begin{equation} \label{PWS}
        \bm{x}_{\text{PWS}} = \mathbf{A}_{\text{PWS}_1} \bm{d}_1 + \mathbf{A}_{\text{PWS}_2} \bm{d}_2,
    \end{equation}
    where $\mathbf{A}_{\text{PWS}_1} = [\bm{u}_1, \dots ,\bm{u}_{32}]$ and$~\mathbf{A}_{\text{PWS}_2}=[\bm{1}_{\mathcal{T}_1}, \bm{1}_{\mathcal{T}_2}, \bm{1}_{\mathcal{T}_3}]$; The vector $\bm{u}_i$ denotes the $i$th eigenvector of the graph Laplacian $\mathbf{L}$, and $\bm{1}_\mathcal{T}$ is the indicator vector of cluster $\mathcal{T}$, i.e., $[\bm{1}_\mathcal{T}]_i=1$ if $i\in\mathcal{T}$ and $[\bm{1}_\mathcal{T}]_i=0$ otherwise. We obtain clusters $\{\mathcal{T}_i\}_{i=1,2,3}$ by performing graph spectral clustering \cite{spectral}. We set $\bm{d}_1 \sim \mathcal{N}(\bm{1}, \mathbf{I})$ and $\bm{d}_2 \sim \mathcal{N}(\bm{0},5\mathbf{I})$.
    Here, the signal subspace is expressed as $\mathbf{A}_{\text{PWS}} = [\mathbf{A}_{\text{PWS}_1},\mathbf{A}_{\text{PWS}_2}]$.
\end{enumerate}

\begin{table*}[tbp]
\centering
\caption{Average reconstruction MSEs in decibels. Bold numbers denote the lowest MSE in each row.
The columns labeled SS represent reconstruction with subspace prior in~\eqref{eq:mx_solution}.
The columns labeled BL is the bandlimited reconstruction, where the numbers are the cutoff frequencies.}
\label{tab:overall_results}
\begin{tabular}{c|c|c|ccccc|ccccc}
\hline
\multicolumn{2}{c|}{Methods} & Prop. & \multicolumn{5}{c|}{SRel} & \multicolumn{5}{c}{SFrob} \\ \hline
\multicolumn{2}{c|}{} & \multirow{2}{*}{SS} & \multirow{2}{*}{SS} & \multicolumn{4}{|c|}{BL} & \multirow{2}{*}{SS} & \multicolumn{4}{|c}{BL} \\ \cline{1-2} \cline{5-8} \cline{10-13}
\multicolumn{2}{c|}{$B$} & & & \multicolumn{1}{|c}{10} & 32 & 100 & 256 & & \multicolumn{1}{|c}{10} & 32 & 100 & 256 \\ \hline \hline
\multirow{2}{*}{HD} & Clean & \textbf{-26.2} & -22.7 & -17.1 & -13.6 & 9.5 & -1.3 & -24.3 & -17.1 & -14.2 & 1.1 & -1.3 \\ \cline{2-2}
& Noisy & \textbf{-25.2} & -22.3 & -17.1 & -12.6 & 32.6 & -1.3 & -23.6 & -17.1 & -13.7 & 28.0 & -1.3 \\ \hline
\multirow{2}{*}{PWS} & Clean & \textbf{-297.2} & -288.2 & -14.8 & 3.7 & -5.4 & -1.2 & -295.2 & -15.2 & -3.2 & -5.3 & -1.2 \\ \cline{2-2}
& Noisy & \textbf{-33.4} & -20.4 & -14.9 & 14.7 & 3.9 & -1.2 & -31.9 & -15.2 & -3.3 & -3.8 & -1.2 \\ \hline
\end{tabular}
\end{table*}

 \subsubsection{Setup}
We partition the nodes into four subsets by cascading the proposed node bipartitioning twice.
The cardinalities of all subsets are therefore $|\mathcal{M}_i|=N/4 = 64,\ i=1,\ldots,4$.
Both noisy and noiseless cases are considered for sampling. 
For the noisy case, additive white Gaussian noise $\bm{\eta}\sim\mathcal{N}(\bm{0},10^{-3}\mathbf{I})$ is added to the signals.

In our algorithm, we experimentally set $L$ in Algorithm~\ref{alg1} and $\beta$ in~\eqref{convex_functions} as $(L,\beta)=(10^3,1)$.
The proposed method is compared with two existing graph partitioning methods: SRel and SFrob~\cite{Comparison}, which are described in Section~\ref{sec:related_work}.

For all partitioning methods, the sampled signals are reconstructed according to~\eqref{eq:mx_solution} with given subspace~$\mathbf{A}_{\text{heat}}$ or $\mathbf{A}_{\text{PWS}}$ to compare the sampling set qualities.

In addition, for SRel and SFrob, we also perform their original reconstruction, i.e., those with the bandlimited assumption.
Specifically, we use  $\mathbf{A}=[\bm{u}_1,\dots,\bm{u}_B]$ for both HD and PWS graph signals.
We experimentally set four bandwidths $B\in \lbrace10,32,100,256\rbrace$.

For all methods, 30 independent runs are performed and the average MSEs are compared.

\begin{figure*}[tbp]
    \centering
    \includegraphics[width=0.820\linewidth]{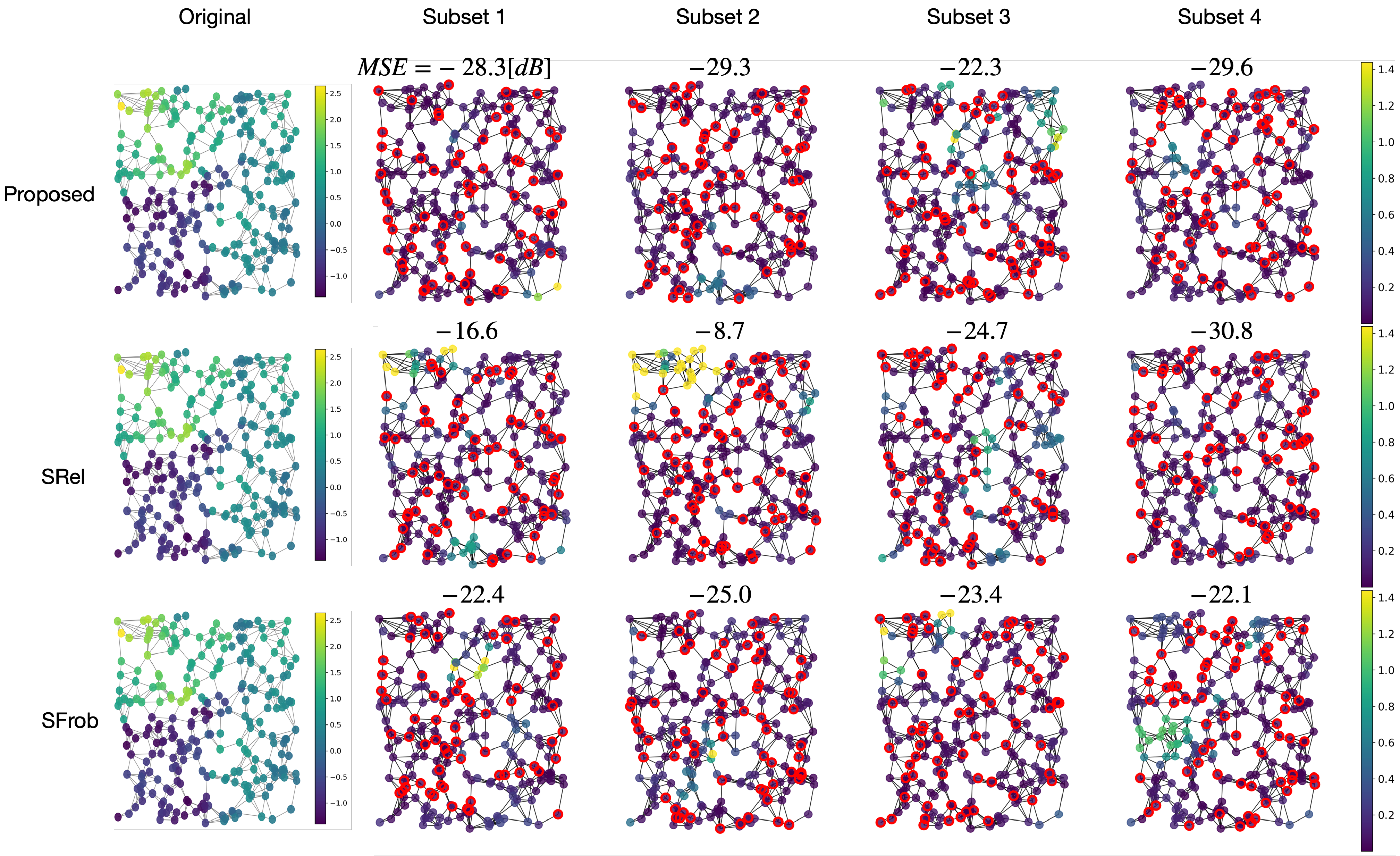}
    \caption{Visualization of the absolute errors between original and reconstructed PWS graph signals. We show the noisy case with reconstruction based on the subspace prior. From top to bottom: The proposed method, SRel, and SFrob. The leftmost column is the original signals (same for all methods). The other columns show the reconstructed signals from sampled subsets. The selected nodes are highlighted by red circles.}
    \label{fig:reconstructed_signals}
\end{figure*}

\subsubsection{Results}

The experimental results are summarized in Table~\ref{tab:overall_results}. 
The proposed method exhibits the lowest MSEs for all cases.
Even when all methods utilize the given subspace for signal reconstruction, the proposed method shows $2$--$5$ dB smaller MSEs than those of SRel and SFrob.
The gain becomes more significant when compared with the original reconstruction methods for SRel and SFrob with the bandlimited assumption.
This is likely because only the proposed method incorporates the signal subspace into graph node partitioning, thereby enabling the selection of sampling subsets that are effective to the signal model.

We also visualize the absolute errors between the original and the reconstructed signals in Fig.~\ref{fig:reconstructed_signals}.
It can be observed that the proposed method presents small reconstruction errors consistently regardless of the subsets. In contrast, SRel sometimes fails to reconstruct the signal as seen in Subset 2, presumably because it does not sample nodes in the upper-left area. Additionally, SFrob has large reconstruction errors in local regions without selected sensors, as in Subset 4.

\subsection{Online Partitioning on Synthetic Graph Signals}
\begin{figure*}[tbp]
    \centering
    \includegraphics[width=0.8\linewidth]{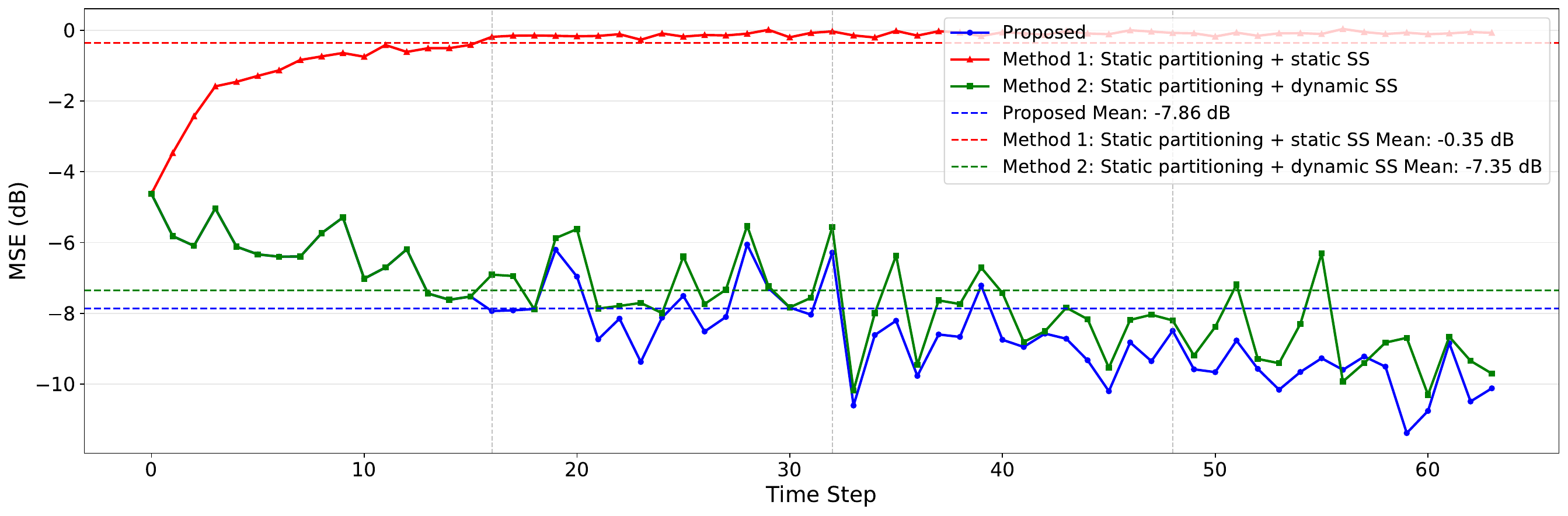}
    \caption{MSE of reconstructed signals~[dB]. The average MSE of each method is plotted as a horizontal dashed line. SS means signal subspace. Vertical dashed lines indicate the time instances when the proposed method updates the partitioning.}
    \label{static_online_experiment}
\end{figure*}
Second, we investigate the effectiveness of the subspace tracking on the performance of sensor scheduling.
For this experiment, we use Algorithm~\ref{alg3} for sensor scheduling and Algorithm~\ref{alg1} for graph node partitioning.

\subsubsection{Graph and Signal Synthesis}
In this experiment, we generate the same graph used in Section~\ref{sec:experiments}-\ref{experiment1}.

As graph signals, we generate time-varying piecewise smooth graph signals based on~\eqref{PWS}. 
In this configuration, $\mathbf{A}_{\text{PWS}_1}$ and $\mathbf{A}_{\text{PWS}_2}$ in~\eqref{PWS} are time-varying and are defined as
\begin{equation}
\begin{split}
    \mathbf{A}_{\text{PWS}_1}(t) &= \mathbf{U} \text{exp}(-\alpha(t)\mathbf{\Lambda})\mathbf{U}^\top,\\ \mathbf{A}_{\text{PWS}_2}(t) &= [\bm{1}_{\mathcal{T}_1(t)},\bm{1}_{\mathcal{T}_2(t)},\bm{1}_{\mathcal{T}_3(t)}],
\end{split}
\end{equation}
where $t$ denotes a time instance and $\alpha(t) = 2 + \frac{1}{8}t$ in which signals become smoother as time passes.
We utilize the same $\bm{d}_1$ and $\bm{d}_2$ as in the previous experiment.

The clusters $\mathcal{T}_1(t),~\mathcal{T}_2(t),~\text{and }\mathcal{T}_3(t)$ are initialized by using spectral clustering~\cite{spectral} at $t=0$.
Subsequently, the cluster memberships of nodes near the boundaries are switched: Nodes located within two hops of the boundaries at $t=0$ are randomly reassigned to one of the other two clusters at each time instance.
Here, $\mathbf{A}_t$ is expressed as $\mathbf{A}_t = [\mathbf{A}_{\text{PWS}_1}(t),\mathbf{A}_{\text{PWS}_2}(t)]$.
We generate the graph signals for the duration of $64$.

\subsubsection{Setup}
The set of nodes is partitioned into 16 subsets with equal size, i.e., $|\mathcal{M}_i| = 16,~ i=1,\ldots ,16$.
The additive white Gaussian noise $\bm{\eta} \sim \mathcal{N}(\bm{0}, 10^{-3}\mathbf{I})$ is added to the signals.
The hyperparameters are set as $(L, \beta) = (10^3, 1)$.
We utilize the minimax recovery~\eqref{eq:mx_solution} for signal reconstruction.

Since there have been no online graph node partitioning methods, we use the following two methods as benchmarks:
\begin{enumerate}
    \item Method 1~(Static partitioning + static signal subspace): In this method, graph node partitioning is fixed, i.e., the initial 16 partitions at $t=0$ is used for all $t$.
    Furthermore, a signal subspace is also fixed to that defined by $\mathbf{A}_0=[\mathbf{A}_{\text{PWS}_1}(0),\mathbf{A}_{\text{PWS}_2}(0)]$ for signal reconstruction.
    \item Method 2~(Static partitioning + dynamic signal subspace): In this method, graph node partitioning is also fixed like Method~1.
    For signal reconstruction, the current signal subspace $\mathbf{A}_t = [\mathbf{A}_{\text{PWS}_1}(t),\mathbf{A}_{\text{PWS}_2}(t)]$ is utilized at each time instance.
\end{enumerate}

For all methods, 10 independent runs are performed and the average MSEs are compared.

\subsubsection{Results}
The results are shown in Fig.~\ref{static_online_experiment}.
As clearly observed, the proposed method and Method 2 are significantly better than Method 1, whose MSE immediately rises and stays high.
This demonstrates that the proposed method successfully tracks the time-varying signal subspace, which is crucial for maintaining high reconstruction accuracy.
Furthermore, the proposed method consistently outperforms Method 2.
This indicates that not only considering the time-varying signal subspace but also employing adaptive partitioning is important for accurate signal reconstruction.

\subsection{Online Partitioning and Dictionary Learning on Real-world Data}
Finally, we evaluate the performance of the proposed online sensor scheduling method based on online graph node partitioning and dictionary learning.
For this experiment, we use Algorithm~\ref{alg2} for sensor scheduling, Algorithm~\ref{alg3} for subspace learning, and Algorithm~\ref{alg1} for graph node partitioning.

\begin{figure}[tbp]
    \centering
    \includegraphics[width=0.96\linewidth]{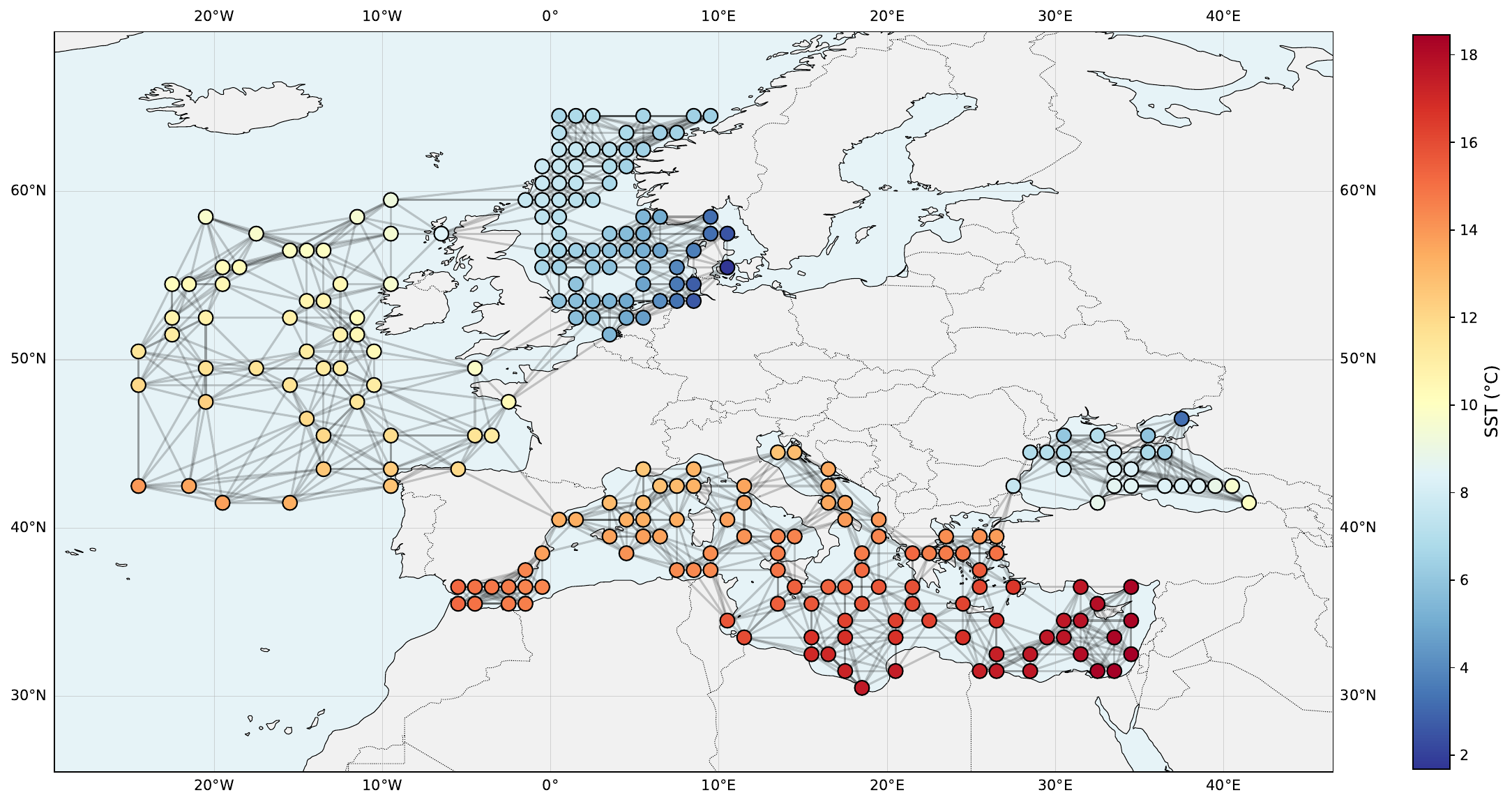}
    \caption{Visualization of a graph signal constructed from global sea surface temperature.}
    \label{real_graph_signal}
\end{figure}

\subsubsection{Setup}
We use the global sea temperature dataset~\cite{seasurfacedata}.
This dataset is composed of snapshots recorded every month from 2016 to 2021.  From sensors all over the world, we randomly select 256 sensors corresponding to the regions of the Mediterranean Sea, the North Sea, Black Sea, and the Northwest Atlantic coast. Subsequently, we create a $k$-NN graph ($k=8$) based on geographical distances between nodes.
We visualize the created graph and the observed signals in Fig.~\ref{real_graph_signal}.

We partition the graph into eight subsets. For the proposed algorithm, parameters are experimentally set as $(L,\beta, D, K)=(10^3,1,20,3\times10^2)$.
The confidence matrix in~\eqref{dictionary_learning}, is defined as:
\begin{equation}
    \mathbf{W}_t = \text{diag}(\bm{m}_t), \label{confidence_matrix}
\end{equation}
where $\bm{m}_t\in\{0,1\}^N$ is the sampling index in~\eqref{m_sampling_operator} for the $t$th instance.
We consider the noisy case where $\bm{\eta}$ in \eqref{sampling} is $\bm{\eta} \sim \mathcal{N}(\bm{0}, 5 \times10^{-1}\mathbf{I})$.

The performance of the proposed method is compared with those of SRel and SFrob~\cite{Comparison}.
In this experiment, we employ a signal reconstruction method in~\eqref{eq:mx_solution} for all approaches to evaluate only the quality of the partitioned sampling subsets.

Furthermore, we conduct an ablation study comparing three dictionary learning configurations to validate the effectiveness of the proposed method.
Specifically, we evaluate the average MSEs under the following configurations:

\begin{enumerate}
    \item Configuration 1: It employs the confidence matrix defined in \eqref{confidence_matrix}. It is identical to the proposed method.

    \item Configuration 2: It assigns uniform weights across the entire reconstructed signal (i.e., $\mathbf{W} = \mathbf{I}$). This setting corresponds to the approach used in conventional dictionary learning methods~\cite{nomurasan}.

    \item Configuration 3: In this configuration, online dictionary learning is performed with the observed signals rather than the reconstructed ones. 
    
    Specifically, the subspace is learned using signals reconstructed via the least-squares reconstruction~\cite{harasan}, i.e., 
    \begin{equation}
        \tilde{\bm{x}}^* = \underset{\tilde{\bm{x}}}{\operatorname{argmin}}~\|\mathbf{S}^\top \tilde{\bm{x}}-\bm{y}\|_2^2 = \mathbf{S}(\mathbf{S}^\top \mathbf{S})^\dagger \bm{y}.
    \end{equation}
    In our setting, this solution performs zero-padding, where non-sampled nodes are set to zero while the observed values are preserved exactly.
    We also utilize the confidence matrix defined by~\eqref{confidence_matrix}.
    Hence, the dictionary learning is performed exclusively on the sampled values.
    
    Note that its performance is evaluated based on signals obtained through the standard reconstruction method~\eqref{eq:mx_solution} with the learned signal subspace.
\end{enumerate}

The performance of these three methods is compared by calculating the average MSE over time.

\subsubsection{Results}
\begin{figure*}[tbp]
    \centering
    \includegraphics[width=0.8\linewidth]{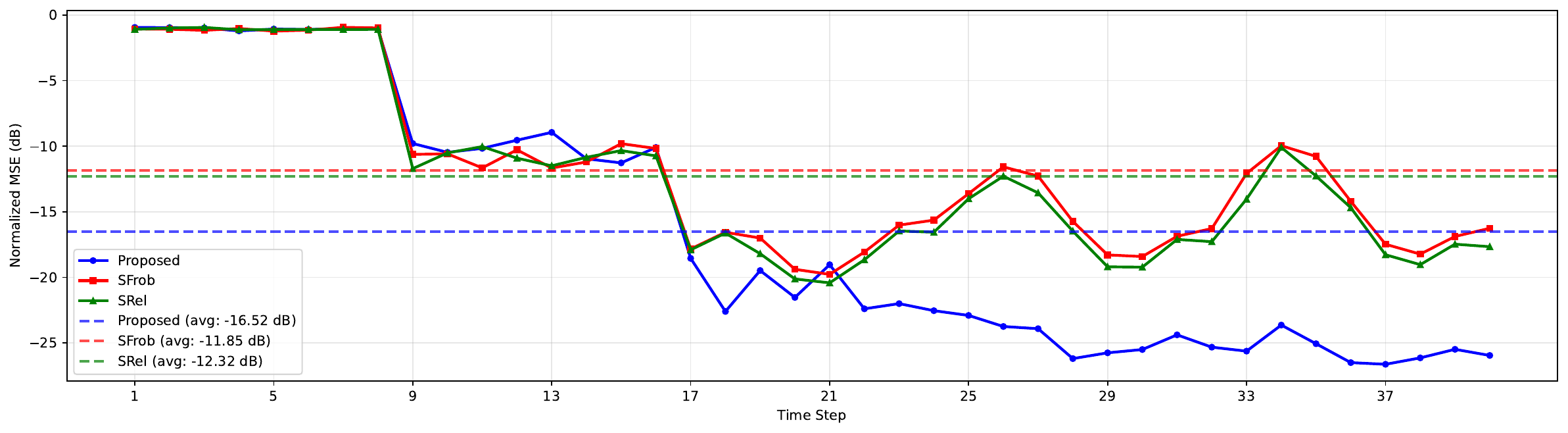}
    \caption{MSE of reconstructed signals~[dB]. The average MSE of each method is plotted as a horizontal dashed line.}
    \label{log_time_series_chart}
\end{figure*}

We visualize the MSEs of the reconstructed signals in Fig.~\ref{log_time_series_chart}. 
The proposed method exhibits the best performance among all methods.
This is because the proposed method incorporates the signal model into its partitioning algorithm, enabling it to achieve a more suitable online partitioning strategy for the learned signal subspace than the other methods.

\begin{figure*}[tbp]
    \centering
    \includegraphics[width=0.85\linewidth]{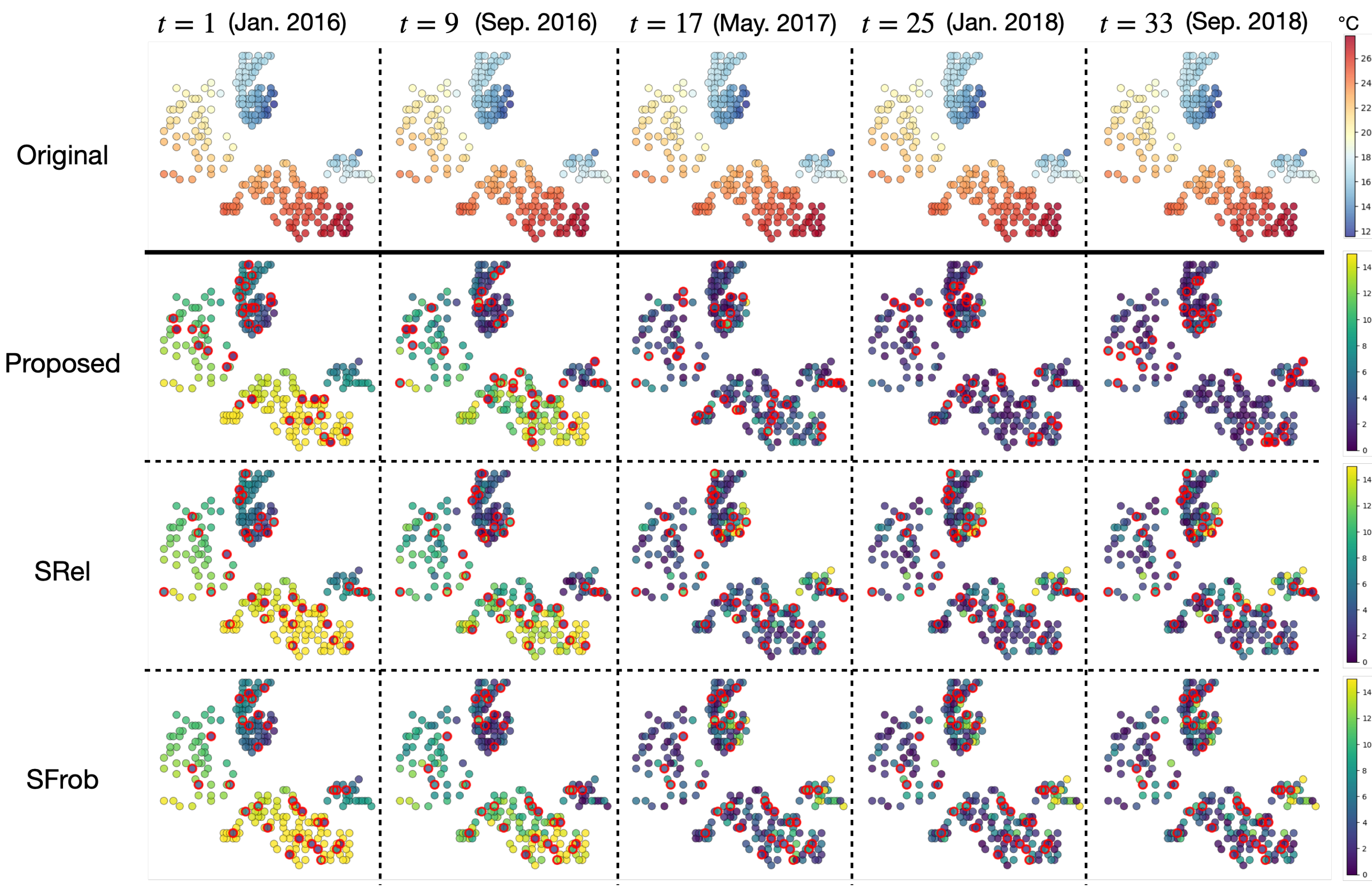}
    \caption{Comparison of the original signal and absolute reconstruction errors at $t = 1, 9, 17, 25, 33$, all obtained using the same subset (Subset 1). The top row displays the original signal, followed by the absolute errors of the proposed method, SRel, and SFrob, respectively. Selected nodes are highlighted with red circles}
    \label{fig:abs_errors_sea_surface}
\end{figure*}

The absolute errors between the original and reconstructed signals are also visualized in Fig.~\ref{fig:abs_errors_sea_surface}.
While SRel and SFrob tend to select nodes that are relatively uniformly distributed across the graph, the proposed method frequently concentrates its selections in specific regions.
This implies that the proposed method can partition nodes into subsets that are suitable for reconstructing full-band graph signals.

\begin{table}[t]
    \centering
    \caption{MSE [dB] of reconstructed signals. Bold numbers denote the lowest MSE.}
    \begin{tabular}{l||ccc}
    \hline 
        Method & Config.~1 & Config.~2 & Config.~3  \\ \hline
         MSE [dB] & $\mathbf{-16.52}$ & $-1.52$ & $-16.31$ \\ \hline
    \end{tabular}
    \label{ablation}
\end{table}

The numerical result of the ablation study is shown in Table~\ref{ablation}.
As expected, Configuration 1 outperforms the Configuration 2.
Configuration 2 could be biased by reconstruction errors in unobserved nodes.
We leave the search for the optimal $\mathbf{W}_t$ beyond~\eqref{confidence_matrix} for future work.
Furthermore, Configuration 1 outperforms the Configuration 3.
This is because Configuration 3 learns the subspace directly from noisy measurements, whereas Configuration 1 utilizes recovered signals for learning, thereby suppressing the influence of observation noise.

\section{CONCLUSION} \label{sec:conclusion}
This paper presents a graph node partitioning method based on graph signal sampling theory.
It partitions nodes into multiple equally-informative subsets that minimize the average reconstruction error.
We formulate the problem as a DC optimization.
We extend the graph node partitioning to the online scenario by introducing dictionary learning for time-varying signal subspace estimation.
Experimental results on synthetic and real-world graphs and graph signals demonstrate that our proposed method outperforms existing graph partitioning methods.

\section*{APPENDIX}
Here, we introduce the derivation of the approximation of~\eqref{eq:bipartition_set} by~\eqref{bipartition2}.
By applying the Neumann series approximation, each term in~\eqref{eq:bipartition_set} is represented as
\begin{equation}
\begin{split} \label{Neumann}
    \text{tr}((\mathbf{S}_i^\top \mathbf{AA}^\top \mathbf{S}_i)^{-1})
     = \frac{1}{\alpha_i}\sum_{n=0}^\infty\text{tr}((\mathbf{I} - \alpha_i\mathbf{S}_i^\top \mathbf{AA}^\top \mathbf{S}_i)^n),\\ 
\end{split}
\end{equation}
where $i=1,2$ and $\alpha_i$ is determined such that it satisfies $\|\mathbf{S}_i^\top \mathbf{AA}^\top \mathbf{S}_i\|_{\text{op}}\leq 1/\alpha$ in terms of the operator norm $\|\cdot\|_{\text{op}}$. We then truncate the approximation up to the second order.
\begin{equation}
\begin{split} \label{Neumann2}
& \frac{1}{\alpha_i}\sum_{n=0}^\infty\text{tr}((\mathbf{I} - \alpha_i\mathbf{S}_i^\top \mathbf{AA}^\top \mathbf{S}_i)^n)\\
    &\approx\frac{1}{\alpha_i}\sum_{n=0}^2\text{tr}((\mathbf{I} - \alpha_i\mathbf{S}_i^\top \mathbf{AA}^\top \mathbf{S}_i)^n)\\ 
    & =  \frac{1}{\alpha_i}\text{tr}(3\mathbf{I} -3\alpha_i\mathbf{S}_i^\top\mathbf{AA}^\top\mathbf{S}_i + (\alpha_i\mathbf{S}_i^\top\mathbf{AA}^\top\mathbf{S}_i)^2).
\end{split}
\end{equation}
Since $\alpha_i$ is sufficiently smaller than unity, the second-order approximation is justified.
Followed by $\mathcal{M}_1 \oplus \mathcal{M}_2 = \mathcal{V}$, we can derive the following relationship:
\begin{equation}
    \begin{split}\label{eq:neumann_const}
    \sum_{i=1}^2\text{tr}(\mathbf{S}_i^\top\mathbf{AA}^\top\mathbf{S}_i)
    &=\sum_{i=1}^2\text{tr}(\mathbf{I}_{\mathcal{M}_i\mathcal{V}}\mathbf{A}^\top\mathbf{A}^\top\mathbf{I}_{\mathcal{M}_i\mathcal{V}}^\top)=\text{tr}(\mathbf{A}\mathbf{A}^\top).
    \end{split}
\end{equation}
Hence, the second term in \eqref{Neumann2} is constant. By ignoring the constant terms, we can approximate \eqref{eq:bipartition_set} by \eqref{bipartition2}.

\bibliography{mybib}

\vfill\pagebreak

\end{document}